\documentstyle[12pt,epsfig]{article}
\begin{document}

\newcommand{\id}{{\hbox{{\rm 1}\kern-.26em\hbox{\rm l}}}}

\begin{center}

{\Large MUTUALLY UNBIASED BASES AND}

\

{\Large HADAMARD MATRICES OF ORDER SIX}

\vspace{10mm}

\hspace{3mm} Ingemar Bengtsson$^{\tiny 1}$ 
\hspace{7mm} Wojciech Bruzda$^{\tiny 2}$ \hspace{11mm} 
\AA sa Ericsson$^{\tiny 1}$ \hspace{8mm}

\

\hspace{4mm} 
Jan-\AA ke Larsson$^{\tiny 3}$ 
\hspace{10mm} Wojciech Tadej$^{\tiny 4}$ 
\hspace{10mm} Karol \.Zyczkowski$^{\tiny 2, 5}$

\vspace{12mm}

{\footnotesize
$^{\tiny 1}${\it Stockholms Universitet, AlbaNova, Fysikum, 
S-106 91 Stockholm, Sweden.}

$^{\tiny 2}${\it Instytut Fizyki im. Smoluchowskiego,
Uniwersytet Jagiello\'{n}ski, ul. Reymonta 4, 30-059 Krak\'{o}w, Poland.}

$^{\tiny 3}${\it Matematiska Institutionen, Link\"opings Universitet, S-581 83 
Link\"oping, Sweden}

$^{\tiny 4}${\it Wydzia\l\ Matematyczno\ -\ Przyrodniczy, Szko\l a 
Nauk \'Scis\l ych, Universytet Kardyna\l a Stefana Wyszy\'nskiego, 
Warszawa, Poland.}

$^{\tiny 5}${\it Centrum Fizyki Teoretycznej, Polska Akademia Nauk,
Al. Lotnik\'{o}w 32/44, 02-668 Warszawa, Poland.}
}
\vspace{8mm}

{\bf Abstract:}

\end{center}

\noindent We report on a search for mutually unbiased bases (MUBs) in 6 dimensions. 
We find only triplets of MUBs, and thus do not come close to the theoretical 
upper bound 7. However, we point out that the natural habitat for sets of 
MUBs is the set of all complex Hadamard matrices of the given order, and we 
introduce a natural notion of distance between bases in Hilbert space. This 
allows us to draw a detailed map of where in the landscape the MUB triplets 
are situated. We use available tools, such as the theory of the discrete 
Fourier transform, to organise our results. Finally we present some 
evidence for the conjecture that there exists a four dimensional family of 
complex Hadamard matrices of order 6. If this conjecture is true the landscape 
in which one may search for MUBs is much larger than previously thought.

\vspace{7mm}

\begin{center}

\noindent {\small ingemar@physto.se, wojtek@gorce.if.uj.edu.pl, asae@physto.se,}

\noindent {\small jalar@mai.liu.se, wtadej@wp.pl, 
karol@tatry.if.uj.edu.pl}

\end{center}

\newpage

{\bf 1. Introduction}

\vspace{5mm} 

\noindent This is a paper on 6 by 6 matrices. Its scope would therefore seem 
rather limited, but we will address two much discussed open problems. The 
first concerns the classification of complex Hadamard 
matrices, and the second the existence (or not) of complete sets of mutually 
unbiased bases. These problems are connected to each other, they are of 
considerable interest in quantum information theory, and they have a long 
history. The study of complex Hadamard matrices was begun by Sylvester 
\cite{Sylvester}, and by Hadamard himself \cite{Hadamard}, while the second problem 
has appeared (in various guises) in quantum theory \cite{Ivanovic, Wootters2}, 
operator algebra theory \cite{Popa}, Lie algebra theory \cite{Kostrikin}, and 
elsewhere \cite{Alltop, Calderbank}. It would seem as if all questions concerning 
single digit dimensions should have been settled by now, but this is not the case.

A complex Hadamard matrix is, by definition, a unitary matrix all of whose 
matrix elements have equal modulus. Much of the mathematical literature 
concerns the special case of real Hadamard matrices, but from now on we will 
take ``Hadamard matrix'' to refer to a complex Hadamard matrix. An obvious 
question concerns the classification of all $N \times N$ Hadamard matrices. 
It has been settled for $N \leq 5$ \cite{Haagerup}, 
but for all larger values of $N$ it is open \cite{Dita, Karol}. 

The columns of a unitary matrix define an orthonormal basis in Hilbert 
space. The basis defined by an Hadamard matrix has the property that the modulus 
of the scalar product between any one of its vectors and any vector in the 
standard basis (whose vectors has only one non-zero entry) is always equal 
to $1/\sqrt{N}$. Again by definition, we say that 
these two bases are mutually unbiased, or MUB for short. One can now ask 
how many bases, with all pairs being MUB with respect to each other, one can 
find in a Hilbert space of a given dimension $N$. It is known that there can be 
at most $N+1$ MUBs, and when $N$ is the power of a prime number the answer is 
$N + 1$ \cite{Wootters2}. For other values of 
$N$ the question is open. For $N = 6$, all that is known is that the number 
of MUBs is at least 3 and at most 7 \cite{Zauner}. 

The set of Hadamard matrices provides a landscape where sets of MUBs live. 
In fact classifying the set of Hadamard matrices is equivalent to classifying 
the set of ordered MUB pairs, up to natural equivalences \cite{KP}. Once such a 
classification has been carried out we proceed to list all MUB triplets, all 
MUB quartets (if any), and so on. However, let 
us admit at the outset that we have carried through this strategy only 
piecemeal---and for $N = 6$ we did not find any MUB quartets. 

In section 2 of this paper we describe the known Hadamard matrices of order 6, 
and sort out some problems concerning equivalences between them. In section 3 
we recall some facts about MUBs, and decide when sets of MUBs should be 
regarded as equivalent. In section 4 we define a useful notion 
of distance between orthonormal bases. The distance attains its maximum when 
the bases are unbiased. In section 5 we describe 
a search for all MUBs whose vectors are composed of rational roots of unity of 
some modest orders; the most interesting of our choices is the 24th root. In 
section 6 we describe the set of all bases that are MUB with respect to both 
the standard and the Fourier bases \cite{Grassl}, and show how this set can 
be understood from properties of the discrete Fourier transform. In section 
7 we describe MUBs of a related type, where the Hadamard matrices have a 
particular block structure. Finally, in section 8 we give some arguments 
suggesting that there exists a four dimensional family of Hadamard 
matrices---the largest known family has two dimensions only. Readers interested 
in Hadamard matrices only can skip sections 3-7, but our main contention is 
that readers interested in MUBs should skip nothing. Section 9 states some 
conclusions.        

{\it Notation}: $z$ is a complex number, and $\bar{z}$ its complex conjugate. 
The rational root of unity $e^{2\pi i/n}$ is denoted 
$\omega$. Its integer powers are called $n$th roots. 
A special case is $q = e^{2\pi i/N}$, where $N$ is the dimension of the 
Hilbert space. A matrix element of the matrix $M$ is 
denoted $M_{ab}$, $M^{\rm T}$ is the transposed matrix, and $M^{\dagger}$ 
is the adjoint. The index $a$ usually runs from $0$ to $5$. 

\vspace{1cm}

{\bf 2. Complex Hadamard matrices}

\vspace{5mm}
 
\noindent According to our definition an {\it Hadamard matrix} $H$ is a unitary 
matrix whose matrix elements obey

\begin{equation} |H_{ab}|^2 = \frac{1}{N} \ , \hspace{8mm} 0 \leq a,b 
\leq N - 1 \ . \end{equation}

\noindent An example that exists for all $N$ is the Fourier matrix ${\bf F}$, 
whose matrix elements are 

\begin{equation} F_{ab} = \frac{1}{\sqrt{N}}q^{ab} \ , \hspace{8mm} 
q \equiv e^{2\pi i/N} \ . \end{equation}

\noindent Two Hadamard matrices $H_1$ and $H_2$ are called {\it equivalent} 
\cite{Haagerup}, written $H_1 
\approx H_2$, if there exist diagonal unitary matrices $D_1$ and $D_2$ and 
permutation matrices $P_1$ and $P_2$ such that 
 
\begin{equation} H_1 = D_1P_1 \cdot H_2 \cdot P_2D_2 \ . \label{equiv} \end{equation} 

\noindent That is to say, we are allowed to rephase and permute rows as well 
as columns. This is an equivalence relation, and we are interested in 
classifying all the equivalence classes for a given matrix size $N$. 

We will usually present our Hadamard matrices in dephased form, which means 
that all elements in the first row and the first column are real and positive. 
This can always be achieved using diagonal unitaries, but it does not fix the 
equivalence class completely. If the 
Hadamard matrix is written in any other form it is said to be enphased.  

For $N = 2$, $3$, and $5$, the Fourier matrix is unique, in the sense that 
it represents the only equivalence class of Hadamard matrices \cite{Haagerup}. 
For $N = 4$ there exists a one parameter family of equivalence classes, 
including the Fourier matrix as well as a real Hadamard matrix \cite{Hadamard}. 
This family exhausts the set of equivalence classes when $N = 4$. 
Let us remark that continuous families 
appear also when $N = 7$ \cite{Karol}, so that their existence does not hinge 
on $N$ not being a prime number.   

We will print concrete Hadamards matrices in boldface. When $N = 6$ the following 
representatives of different equivalence classes are known to us: 

\begin{itemize}

\item{A two parameter family ${\bf F}(x_1,x_2)$, including the Fourier matrix 
${\bf F}(0,0)$.}

\item{The transpose ${\bf F}^{\rm T}(x_1,x_2)$ of the above.} 

\item{A circulant matrix ${\bf C}$ found by Bj\"orck \cite{Bjorck}, and 
its complex conjugate.}

\item{A one parameter family ${\bf D}(x)$, including a matrix ${\bf D}$ composed 
of fourth roots of unity, called Di\c{t}\u{a}'s matrix \cite{Dita}.} 

\item{A one parameter family ${\bf B}(\theta)$ that interpolates between 
${\bf C}$ and ${\bf D}$ \cite{Nicoara}.}

\item{Tao's matrix ${\bf S}$, composed of third roots of unity \cite{Moorhouse, Tao}.}

\end{itemize}

\noindent Attribution may be difficult; the Di\c{t}\u{a} family, or 
representatives thereof, was discovered several times \cite{Ufnarovski, Haagerup, 
Zauner}. With one exception, namely the family ${\bf B}(\theta)$, the known 
continuous families are of the special kind called affine families 
\cite{Karol}, which means that some of the matrix elements can be multiplied 
with free phase factors in such a way that the matrix remains Hadamard.  

We now go through the items in our list. The {\it Fourier family} is explicitly 

\begin{equation} {\bf F}(x_1,x_2) = 
\left[ \begin{array}{cccccc} 1 & 1 & 1 & 1 & 1 & 1 \\ 
1 & qz_1 & q^2z_2 & q^3 & q^4z_1 & q^5z_2 \\
1 & q^2 & q^4 & 1 & q^2 & q^4 \\
1 & q^3z_1 & z_2 & q^3 & z_1 & q^3z_2 \\
1 & q^4 & q^2 & 1 & q^4 & q^2 \\
1 & q^5z_1 & q^4z_2 & q^3 & q^2z_1 & qz_2 
\end{array}\right] \ , \hspace{5mm} 
\begin{array}{l} z_1 \equiv e^{2\pi ix_1} \\ 
\ \\ \ \\ z_2 \equiv e^{2\pi ix_2} \end{array} \ . \label{Fourier} 
\end{equation}

\noindent The two free parameters arise essentially because a six dimensional 
space can be written as a tensor product; see section 7. This is an affine 
family, since there are no restrictions on the phase factors $z_1$ and $z_2$. 

\begin{figure}[htp]
        \centerline{ \hbox{
                \epsfig{figure=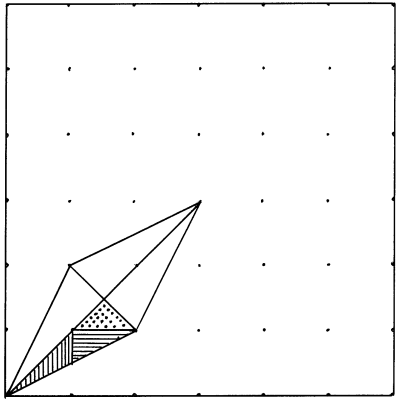,width=55mm}}}
        \caption{\small The affine family ${\bf F}(x_1,x_2)$ is divided 
into 144 equivalent triangles of equal area, as described in the text. The 
lattice shown in the background are the 6th roots.}
        \label{fig:1}
\end{figure} 

There are equivalence relations connecting different matrices within this 
family. They form a discrete group, and thus they will affect the size but not 
the dimensionality of the affine family. Let us think of the 
parameter space as a square in the $(x_1, x_2)$ plane (see fig.~\ref{fig:1}). 
By inspection one finds cyclic translation groups of order 6:   

\begin{eqnarray} {\bf F}(x_1,x_2) \approx {\bf F}(x_1+2/6,x_2+1/6) 
\approx {\bf F}(x_1+4/6,x_2+2/6)
\approx {\bf F}(x_1,-x_2) , \nonumber \\
\ \\
{\bf F}(x_1,x_2) \approx {\bf F}(x_1+1/6,x_2+2/6) \approx {\bf F}(x_1+2/6,x_2+4/6)
\approx {\bf F}(-x_1,x_2) . \nonumber \end{eqnarray}

\noindent They divide the square into $12$ copies 
of a fundamental region that is a parallellogram. 
There are further equivalences corresponding to transformations of order 
$2$. By inspection

\begin{equation} {\bf F}(x_1,x_2) \approx {\bf F}(x_2,x_1) \ . \end{equation}

\noindent If we permute the rows and then dephase the row that 
ends up in the first place, we find that 

\begin{equation} {\bf F}(x_1,x_2) \approx {\bf F}(-x_1,-x_2) \ . \end{equation}

%
%

\noindent As a result, our parallellogram will be divided into four 
equivalent triangles. Finally we can permute the columns and dephase the  
column that ends up in the first place. This gives a final cyclic 
equivalence group of order 3:

\begin{equation} {\bf F}(x_1,x_2) \approx {\bf F}(x_2-x_1, -x_1) \approx 
{\bf F}(-x_2, x_1-x_2) \end{equation}

\noindent Considered as transformations of our square, these last 
transformations are not isometries but they do preserve area. Hence the 
triangles that we had in the previous step are themselves divided 
into three triangles of equal area, one of which is our final fundamental region 
of inequivalent Hadamard matrices. A convenient choice of fundamental region 
is a triangle with corners at $(0,0)$, $(1/6,0)$ and $(1/6, 1/12)$; the 
original square is covered by $144$ copies of this region. The family 
${\bf F}^{\rm T}(x_1,x_2)$ works (at this stage) in an entirely analogous way. 

{\it Bj\"orck's circulant Hadamard matrix} is 

\begin{equation} {\bf C} = \left[ \begin{array}{cccccc} 
1 & id & - d & - i & - \bar{d} & i\bar{d} \\
i\bar{d} & 1 & id & - d & - i & - \bar{d} \\
- \bar{d} & i\bar{d} & 1 & id & - d & - i \\
- i & - \bar{d} & i\bar{d} & 1 & id & - d \\
- d & - i & - \bar{d} & i\bar{d} & 1 & id \\
id & - d & - i & - \bar{d} & i\bar{d} & 1 \end{array} \right] \ , \hspace{6mm} 
\bar{d}d = 1 \ .  
\label{Bjorck} \end{equation}

\noindent See section 6 for further discussion of circulant Hadamard matrices. 
The complex number $d$ has modulus unity and is 

\begin{equation} d = \frac{1-\sqrt{3}}{2} + i\sqrt{\frac{\sqrt{3}}{2}} 
\hspace{5mm} \Rightarrow \hspace{5mm} 
d^2 - (1-\sqrt{3})d + 1 = 0 \ . \label{d} \end{equation}

\noindent No affine family stems from this matrix \cite{Karol}. We will have 
more to say about this matrix later; for now let us just mention that all 
Hadamard matrices equivalent to a circulant matrix have 
been listed by Bj\"orck and coauthors \cite{Bjorck}, for $N \leq 8$. 

The {\it Di\c{t}\u{a} family} is 

\begin{equation} {\bf D}(x) = \frac{1}{\sqrt{6}}\left[ \begin{array}{cccccc} 
1 & 1 & 1 & 1 & 1 & 1 \\
1 & - 1 & i & - i & - i & i \\
1 & i & - 1 & iz & - iz & - i \\
1 & - i & i\bar{z} & - 1 & i & - i\bar{z} \\ 
1 & - i & - i\bar{z} & i & - 1 & i\bar{z} \\
1 & i & - i & - iz & iz & - 1 \end{array} 
\right] \ , \hspace{6mm} z \equiv e^{2\pi ix} \ . \label{Dita2} \end{equation}

\noindent There are equivalence relations within this family too. In fact 

\begin{equation} 
{\bf D}(x) \approx {\bf D}(x + 1/2) \approx {\bf D}(-x + 1/4) \ . \end{equation}

\noindent Hence we can take $- 1/8 \leq x \leq 1/8$ without loss of 
generality. The matrices ${\bf D}(x)$ can be written on block circulant form; 
see section 7. 

The {\it Hermitian family} ${\bf B}(\theta)$ was found very recently, by 
Beauchamp and Nicoara \cite{Nicoara}, and consists---up to equivalences---of 
all Hermitian Hadamard matrices of order 6. Explicitly it is 

\begin{equation} {\bf B}(\theta) = \frac{1}{\sqrt{6}}\left[ \begin{array}{cccccc} 
1 & 1 & 1 & 1 & 1 & 1 \\
1 & - 1 & - \bar{x} & - y & y & \bar{x} \\
1 & -x & 1 & y & \bar{z} & - \bar{t} \\
1 & - \bar{y} & \bar{y} & -1 & -\bar{t} & \bar{t} \\
1 & \bar{y} & z & -t & 1 & -\bar{x} \\
1 & x & -t & t & -x & -1 \end{array} \right] 
\end{equation}

\noindent where $(x,y,z,t)$ are complex numbers of modulus one, related by 

\begin{equation} y = e^{i\theta} \hspace{2cm} t = xyz \end{equation}

\begin{equation} z = \frac{1+2y-y^2}{y(-1+2y+y^2)} \end{equation}

\begin{equation} x = \frac{1+2y +y^2 \pm \sqrt{2}\sqrt{1+2y+2y^3+y^4}}
{1+2y-y^2} \ , \end{equation}

\noindent with $\theta$ providing a free parameter. The two branches 
of the square root lead to equivalent families. This is an 
example of a non-affine family (and the only such example known to us for 
$N = 6$).  

The phase $\theta$ cannot be chosen arbitrarily; an interval around 
$\theta = 0$ is excluded. The Hadamard property requires that   

\begin{equation} |y|^2 = 1 \hspace{5mm} \Rightarrow \hspace{5mm} 
|x|^2 = |z|^2 = |t|^2 = 1 \ . 
\end{equation} 

\noindent On examining this point one sees that this restricts the allowed 
values of $y = \cos{\theta} + i\sin{\theta}$. There are no difficulties with $z$, 
but the numerator of $x$ becomes 

\begin{equation} 1+2y +y^2 \pm \sqrt{2}\sqrt{1+2y+2y^3+y^4} = 2y(1 + \cos{\theta} 
\pm\sqrt{2\cos{\theta} + 2\cos^2{\theta} - 1}) . \end{equation}

\noindent The absolute value of this expression hinges crucially on the sign 
inside the square root; the absolute value of $x$ will be equal to one if 
that sign is negative. This restricts the phase to 

\begin{equation} \cos{\theta} \leq \frac{\sqrt{3} - 1}{2} \ . \end{equation}

\noindent The allowed range of $\theta$ has end points, with vanishing square root, 
at $y = -\bar{d}$ and $y = - d$. The curve is not closed; its end points 
correspond to permutations 
of Bj\"orck's matrix ${\bf C}$ and its complex conjugate. This is also the case if 
$y = \bar{d}^2$ or $y = d^2$. If $y = - 1$ or $y = \pm i$ we obtain permutations 
of Di\c{t}\u{a}'s matrix.  

{\it Tao's matrix} never appeared in our searches for MUBs, so we do not give it 
explicitly here. Since Tao's matrix is composed of 3d roots only, one may ask 
whether there are any $N = 6$ Hadamard matrices based only on 5th or 7th roots. 
Using a computer program to be described in section 5 we 
have shown that no such Hadamard matrix exists. 

\vspace{1cm}

{\bf 3. Preliminaries on MUBs}

\vspace{5mm}

\noindent We now turn to {\it Mutually Unbiased Bases}, or MUBs. First we recall 
some definitions. Fix an orthonormal basis $|e_a\rangle$ in an $N$ 
dimensional Hilbert space. A unit vector $|f\rangle$ is said to be unbiased 
with respect to the fixed basis if for all $a$ 

\begin{equation} |\langle e_a|f\rangle |^2 = \frac{1}{N} \ . \end{equation} 

\noindent The important thing is that the right hand side is constant; its 
value is a consequence. A pair of orthonormal bases is said to be 
an unbiased pair if all the vectors in one of the bases are unbiased with 
respect to the other. One can go on to define larger sets of mutually 
unbiased bases, or MUBs, in the obvious way. It is known 
that the number of MUBs one can find is bounded from above by $N+1$, but 
it is not known if this bound can be achieved unless $N$ is a power of 
a prime number. A set of $N+1$ MUBs, if it exists, is 
known as a complete set. 

If one member of a set of MUBs is represented by the columns of the unit 
matrix, all the other bases must be represented by the columns of a set of Hadamard 
matrices. A complete set of $N+1$ MUBs exists if we can find $N$ enphased Hadamard 
matrices representing bases that are MUB with respect to each other. Altogether 
then we have $N$ Hadamard matrices that are 
said to be {\it Mutually Unbiased Hadamards}, or MUHs \cite{Karol}. Note that two 
Hadamard matrices $H_1, H_2$ are unbiased if and only if 

\begin{equation} H^{\dagger}_1H_2 = H_3 \ , \end{equation}

\noindent where $H_3$ is an Hadamard matrix too. We say that a set of MUHs all 
of whose members are equivalent in the sense of section 2 is {\it homogeneous}, 
otherwise it is {\it heterogeneous}. The standard 
construction of complete sets of MUBs in prime power dimensions \cite{Wootters2} 
gives a homogeneous set of $N$ MUHs. 

When are two pairs of MUBs equivalent to each other? Let $(M_0, M_1)$ denote 
an ordered pair of MUBs, and $\{ M_0, M_1 \}$ an unordered pair, with each basis 
represented as the columns of a unitary matrix. We are not interested in the 
order of the constituent vectors, nor in the phase factors multiplying these 
vectors. Hence, with $P$ a permutation and $D$ a diagonal unitary matrix, 

\begin{equation} (M_0PD, M_1P'D') \approx (M_0,M_1) \ . \end{equation}

\noindent We also declare that an overall unitary transformation of the Hilbert 
space is irrelevant, so that  

\begin{equation} (UM_0, UM_1) \approx (M_0, M_1) \ . \label{Uequiv} 
\end{equation}

\noindent Then the pair can always be represented in the form $({\id}, H)$, 
where $H$ is an Hadamard matrix. It follows 
that the equivalence relation (\ref{equiv}), used for Hadamard matrices, 
is the correct equivalence relation also for ordered MUB pairs; 

\begin{equation}  ({\id}, H_1) \approx ({\id}, H_2) \hspace{5mm} 
\Leftrightarrow \hspace{5mm} H_1 \approx H_2 \ . \end{equation}

\noindent From eq. (\ref{Uequiv}) we know that $({\id}, H) \approx 
(H^\dagger , {\id})$. For unordered MUB pairs this means that 

\begin{equation} \{ {\id}, H\} \approx \{ {\id}, H^\dagger \} \ . \end{equation}

\noindent Therefore, the {\it condition for unordered pairs to be equivalent} is 

\begin{equation} \{ {\id}, H_1\} \approx \{ {\id}, H_2\} \hspace{5mm} 
\Leftrightarrow \hspace{5mm} \begin{array}{ll} 
\mbox{either} \ H_1 \approx H_2 & \ \\ \ & \ \\ 
\mbox{or} \ H_1 \approx H_2^\dagger & \ . \end{array} \end{equation}

\noindent The discussion can be extended to larger sets of MUBs, ordered or 
unordered ---although the equivalences will become harder to check as the 
number of MUBs increases.

For the Hadamard matrices we know, it is true that 

\begin{equation} [{\bf F}(x_1, x_2)]^\dagger \approx 
{\bf F}^{\rm T}(x_1,x_2) \label{FTF} \end{equation}

\begin{equation} [{\bf D}(x)]^{\dagger} \approx {\bf D}(-x) \end{equation} 

\begin{equation} {\bf C}^\dagger \approx {\bf C} \hspace{12mm} 
{\bf S}^\dagger \approx {\bf S} \ , \end{equation}

\noindent and finally the Beauchamp-Nicoara family is Hermitian by 
construction. Thus the set of unordered MUB pairs is smaller 
than the set of inequivalent Hadamard matrices. In particular, $\{ {\id}, 
{\bf F}(x_1,x_2)\}$ and $\{ {\id}, {\bf F}^{\rm T}(x_1,x_2)\}$ are equivalent 
when considered as unordered pairs.
 
Let us return to the MUBs themselves. It is useful to think of them as sets 
of density matrices, rather than as sets of vectors in Hilbert space. Recall 
that a unit vector 
$|e\rangle $ in an $N$ dimensional Hilbert space corresponds to a projector 
$|e\rangle \langle e|$, which is an Hermitian matrix of trace unity, and as 
such can be regarded as a vector in an $N^2 - 1$ real dimensional vector 
space whose elements are matrices. The origin of this vector space is 
naturally chosen to sit at the matrix $\rho_* = \frac{1}{N}{\id}$, and 
then its vectors are traceless matrices. Explicitly the correspondence is  

\begin{equation} |e\rangle \hspace{5mm} \rightarrow \hspace{5mm} {\bf e} = 
\sqrt{\frac{2N}{N-1}}\left( |e\rangle \langle e| - \rho_\star \right) 
\ . \label{correspondence} \end{equation}

\noindent The vector ${\bf e}$ is a unit vector with respect to the scalar product 

\begin{equation} {\bf e}\cdot {\bf f} = \frac{1}{2}\mbox{Tr}\ {\bf e}
{\bf f} \ . \end{equation}

\noindent The distance between two matrices $A$ and $B$ is given by 

\begin{equation} d^2(A,B) = \frac{1}{2}\mbox{Tr}(A-B)^2 \ . \end{equation} 

\noindent This is the Euclidean Hilbert-Schmidt distance.

It is important to realize that although any unit vector in Hilbert space 
gives rise to a unit vector in the larger space, it is only a small subset 
of the latter that can be realized in this way. The convex cover of this 
small subset is the set of all density matrices (or quantum states).   
An orthonormal basis in the Hilbert space corresponds to $N$ vectors that 
form a regular simplex in the larger space. This simplex spans an $N-1$ 
dimensional plane through the origin. It is moreover easy to see that 

\begin{equation} |\langle e|f\rangle |^2 = \frac{1}{N} \hspace{5mm} 
\Leftrightarrow \hspace{5mm} {\bf e}\cdot {\bf f} = 0 \ . \end{equation}

\noindent Therefore, if the unit vectors belong to a pair of MUBs, the 
corresponding vectors in the large vector space are orthogonal. It 
follows that a pair of MUBs span two totally orthogonal $(N-1)$-planes 
through the origin. This fact is at the bottom of the reason why 
MUBs are useful in quantum state tomography \cite{Wootters2}. And we 
see immediately that there can be at most $N+1$ MUBs, since there is 
no room for more than $N+1$ totally orthogonal $(N-1)$-planes in an 
$N^2-1$ dimensional space.  

\vspace{1cm}

{\bf 4. Interlude: a distance between bases}

\vspace{5mm}

\noindent We need a distance between bases in Hilbert space, such that 
it becomes maximal if the bases are MUB. It should be natural and easy 
to compute, but we do not insist on any precise operational meaning 
for it. 

As explained in section 3, a basis in an $N$ dimensional Hilbert space 
spans an $(N-1)$-plane in a real vector space of $N^2 - 1$ dimensions. There 
is a standard way to define a distance between such planes, namely to regard 
them as points in a Grassmannian, in itself embedded in the surface of 
a sphere within an Euclidean space of 
a quite high dimension \cite{BZ}. The procedure 
is analogous to the one we followed when we transformed our Hilbert space 
vectors into points in a space embedded within the space of traceless 
matrices. The details are as follows. Starting 
from a basis $|e_a\rangle $ in Hilbert space, use the recipe given in eq. 
(\ref{correspondence}) to form the $N$ vectors ${\bf e}_a$.  Expand these 
vectors relative to a basis. Then form the $(N^2 - 1)\times N$ matrix  

\begin{equation} B = \sqrt{\frac{N-1}{N}}
[ {\bf e}_1 \ {\bf e}_2 \ \dots \ {\bf e}_N] 
\ . \end{equation}

\noindent It has rank $N-1$. Next introduce an $(N^2 -1)\times (N^2 -1)$ 
matrix of trace $N-1$, projecting onto the $(N-1)$ dimensional 
plane spanned by the ${\bf e}_a$:  

\begin{equation} P = B \ B^{\rm T} = \frac{N-1}{N}[ {\bf e}_1 
\ \dots \ {\bf e}_N]\left[ \begin{array}{c} {\bf e}^{\rm T}_1 \\ 
\dots \\ {\bf e}_N^{\rm T}\end{array} \right] \ . \end{equation}

\noindent It is easy to check (through acting on ${\bf e}_a$ say) that 
this really is a projector. The {\it chordal Grassmannian distance} between 
two $(N-1)$-planes is defined in terms of the corresponding projectors as 

\begin{equation} D^2_c(P_1, P_2) = \frac{1}{2(N-1)}\mbox{Tr}(P_1 - P_2)^2 = 
1 - \frac{1}{N-1}\mbox{Tr}P_1P_2 \ . \end{equation}

\noindent There is an analogy to how the density 
matrices were defined in the first place, and to the Hilbert-Schmidt distance 
between them. 

We can think of the projectors $P$ as points on the surface of a sphere 
in ${\bf R}^M$, where---as it happens--- 

\begin{equation} M = \frac{N^4 - N^2 - 2}{2} \ . \end{equation}

\noindent This explains the name ``chordal distance''. It is however important 
to realize that the Grassmannian of $(N-1)$-planes forms a very small subset 
of this sphere. Its dimension is $N(N-1)^2$. And then bases in Hilbert 
space correspond to a small subset of the Grassmannian. 

The chordal distance has been used in studies of packing problems 
for planes and subspaces \cite{Conway}. It has a number of advantages 
when compared to more sophisticated distances, such as the geodesic 
distance within the Grassmannian (which would be analogous to the 
Fubini-Study distance between pure quantum states). It is useful to know 
that the chordal distance can be written as a function of the principal angles 
$\theta_i$, namely

\begin{equation} D^2_c = 1 - \frac{1}{N-1}\sum_{i = 1}^{N-1}\cos^2{\theta_i} = 
\frac{1}{N-1}\sum_{i=1}^{N-1}\sin^2{\theta_i} \ . \end{equation} 

\noindent Here the first principal angle is defined as 

\begin{equation} \cos{\theta_1} = \max {\bf u}_1\cdot {\bf v}_1 \ , 
\end{equation}

\noindent where ${\bf u}_1$ and ${\bf v}_1$ are unit vectors belonging 
to the respective $(N-1)$-planes, chosen so that their scalar product 
is maximized. The second principal angle is defined 
by maximizing the scalar product between unit vectors in the orthogonal 
complements to ${\bf u}_1$ and ${\bf v}_1$, and so on. It is then 
clear that 

\begin{equation} 0 \leq D^2_c \leq 1 - \frac{k}{N-1} \ , \label{tomo} \end{equation}

\noindent where $k$ is the dimension of the intersection of the two 
planes. The distance is maximal if and only if the $(N-1)$-planes are totally 
orthogonal. 

Now consider two $(N-1)$-planes spanned by vectors corresponding to two 
bases $|e_a\rangle $ and $|f_a\rangle $ in Hilbert space. Working through 
the details, one finds that the distance squared between the bases is 

\begin{equation} D_c^2(P_1, P_2) = 1 - \frac{1}{N-1}\sum_{a=0}^{N-1}
\sum_{b=0}^{N-1}\left( 
|\langle e_a|f_b\rangle |^2 - \frac{1}{N}\right)^2 \ . \end{equation}

\noindent Thus, between bases in Hilbert space,  

\begin{equation} 0 \leq D^2_c \leq D_{\rm max}^2 = 1 \ . \end{equation} 

\noindent The distance attains its maximum value if and only if the 
bases are MUB. A set of MUBs forms a regular equatorial simplex on 
the sphere in ${\bf R}^M$, although there will be many regular equatorial 
simplices that do not arise in this way. 

What is $D^2_c$ on the average, for two bases picked at random in Hilbert space? 
To answer this question we represent one basis by the standard basis. 
Then we choose a vector at random according to the Fubini-Study 
measure, choose a vector in its orthogonal complement again according to the 
Fubini-Study measure (in one dimension lower), and so on until we have a 
complete basis. The resulting measure is a measure on the flag manifold 
$U(N)/[U(1)]^N$. In practice the calculation is simple; the average is 

\begin{equation} \left< D^2_c\right> = \left< 1 - \frac{1}{N-1}\sum_a\sum_b \left( 
|\langle e_a|f_b\rangle |^2 - \frac{1}{N}\right)^2 \right> \ . \end{equation}

\noindent Using the linearity of the average, together with the fact that 
the $N^2$ terms in the sum must have equal averages, we can rewrite this as 
 
\begin{equation} \left< D^2_c \right> = 1 - \frac{N^2}{N-1} \left< 
\left( |\langle e_0|
f_0\rangle |^2 - \frac{1}{N}\right)^2 \right> \ , \end{equation}

\noindent Hence it is enough to calculate the average of a function of the 
modulus of a component of a random vector, using the Fubini-Study measure. 
How to do this is described elsewhere \cite{Mello,BZ}; the answer we arrive at is  

\begin{equation} \left< D^2_c\right> = \frac{N}{N+1} 
\ . \end{equation}


\noindent For $N = 6$ we have $\left< D^2_c\right> = 0.86$; as $N$ grows 
the average distance squared approaches the maximum value $1$. Note that 
the average depends smoothly on $N$, whatever the maximal number of MUBs 
in $N$ dimensions may be. 

One can ask for a function of $N+1$ points on the sphere in ${\bf R}^M$, 
whose maximum is attained when the points form a regular equatorial simplex. 
Given that $N+1 \leq M$ it happens that 

\begin{equation} f = \sum_{i= 1}^{N+1}\sum_{j=1}^{N+1}D^2_c(P_i, P_j) 
\ , \end{equation}

\noindent is such a function. 
To see this, use the Euclidean norm on the space in which the 
Grassmannian of $(N-1)$-planes is embedded, and normalise it so that 
the projectors correspond to $N+1$ unit vectors ${\bf E}_i$. Then 

\begin{equation} \sum_{i<j}||{\bf E}_i - {\bf E}_j||^2 
= N(N+1) - 2\sum_{i<j}{\bf E}_i\cdot {\bf E}_j \ . \end{equation}

\noindent On the other hand 

\begin{equation} N+1 + 2\sum_{i<j}{\bf E}_i\cdot {\bf E}_j = 
||\sum_i{\bf E}_i||^2 \geq 0 \ . \label{inequal} \end{equation} 

\noindent A minimum of the last expression corresponds to a maximum 
of the function $f$. A regular equatorial simplex clearly saturates the 
inquality (\ref{inequal}), given that $N+1$ is smaller than the dimension 
of the space that we are in. This proves our point. There are 
other configurations besides the desired one that also saturate the bound. 
Since they do not arise from bases in the underlying Hilbert space we 
can ignore such configurations. Still we do not claim that $f$ is the most 
useful function of its kind. 

Equipped with our notion of distance (so that we can quantify 
our failures), we can look for complete sets of 7 MUBs in ${\bf C}^6$. The 
natural way to proceed, given the results of this section, is to maximise 
the function 

\begin{equation} f = \sum_{i=1}^{7}\sum_{j=1}^{7}D^2_c(P_i, P_j) = 
\sum_{i=1}^{7}\sum_{j=1}^{7}\left( 1 - \frac{1}{5}\sum_{a=0}^5\sum_{b = 0}^5
\left( |\langle e^{(i)}_a|e^{(j)}_b\rangle |^2 - \frac{1}{N}\right)^2 
\right) \ , \label{F} \end{equation}

\noindent with the understanding that the $|e^{(i)}_a\rangle $ are orthonormal 
bases in Hilbert space. There is no {\it a priori} reason to believe that the 
upper bound on $f$ can be attained, because we are confined to a small subset 
of all possible $(N-1)$-planes. Still, it might be interesting to find the 
``best'' solution in this sense. A similar but more sophisticated procedure 
has been successfully used to find a special kind of overcomplete bases known 
as SIC-POVMs \cite{Renes}. We have not attempted such a calculation however. 

\vspace{1cm}

{\bf 5. MUBs composed from rational roots of unity}

\vspace{5mm}

\noindent In our search for MUBs we rely on the classification of 
ordered MUB pairs through Hadamard matrices. 
Without loss of generality we assume that the first MUB is represented by the 
standard basis ${\id}$, and call it $(\mbox{MUB})_0$. 
Then we choose a representative of an equivalence class of Hadamard matrices, 
and call it $(\mbox{MUB})_1$. Assume it to be given in dephased form. 
All bases that are MUB with respect to the first two must then be 
represented by enphased Hadamard matrices, and we proceed to look for them.   

For $N \leq 5$ there exist maximal sets of $N + 1$ MUBs, 
and these sets are unique up to an overall unitary transformation. Given 
that $(\mbox{MUB})_1$ 
is not unique for $N = 4$, this is perhaps a little surprising. But for 
$N = 4$ it is possible to find inequivalent and incomplete sets of MUBs, 
that cannot be extended to complete sets \cite{Beth}. To get the complete 
set one must start with the real Hadamard matrix, rather than the Fourier 
matrix, as $(\mbox{MUB})_1$. Given that Hadamard matrices of order 6 are 
highly non-unique, the question where to place $(\mbox{MUB})_1$ becomes 
non-trivial for $N = 6$. 

All known complete sets of MUBs are built from vectors all 
of whose components are $N$th or $2N$th roots of unity, depending on whether 
the dimension $N$ is odd or even \cite{Wootters2, Klappenecker}. 
We therefore made a program that lists, for $N = 6$, all orthonormal bases 
whose vectors are composed of 12th roots. They are candidates for $(\mbox{MUB})_1$. 
Next, for each $(\mbox{MUB})_1$ the program lists the set of all vectors 
composed from 12th roots and unbiased with respect to it. The program also 
lists all orthonormal bases that can be constructed from each set of unbiased 
vectors. They are candidates for $(\mbox{MUB})_2$. Finally we prune the list 
so that it contains only inequivalent Hadamard matrices as $(\mbox{MUB})_1$, 
and if there are several $(\mbox{MUB})_2$ we compute the distance between 
them to see if we get any sets of four MUBs in this way. 
Note that this means that the part of the landscape of Hadamard matrices 
we search in consists of the Fourier and Di\c{t}\u{a} families, the Tao 
matrix, enphased versions of these, and (probably) nothing more. The 
analogous calculation when $N$ is a power of a prime would have found all 
known complete sets of MUBs, but for $N = 6$ only triplets of MUBs turned up. 

The 12th roots list we arrived at is   

\

${\bf F}(0,0)$ admits 4 candidates for $(\mbox{MUB})_2$. 

${\bf F}(1/6,0)$ admits 1 candidate for $(\mbox{MUB})_2$.

${\bf F}^{\rm T}(1/6,0)$ admits 1 candidate for $(\mbox{MUB})_2$.

\

\noindent This is all. The 4 bases that are MUB with respect to ${\bf F}(0,0)$ 
will be discussed in the next section, the basis that is MUB with respect to 
${\bf F}^{\rm T}(1/6,0)$ is an enphased version of ${\bf F}(0,0)$, while the 
basis that is MUB with respect to ${\bf F}(1/6,0)$ is an enphased version of 
itself. 

We redid the entire calculation using 24th roots. This 
resulted in the following additional entries in the list:

\

${\bf F}(1/6, 1/12)$ admits 4 candidates for $(\mbox{MUB})_2$.

${\bf D}(1/8)$ admits 4 candidates for $(\mbox{MUB})_2$. 
 
\

\begin{figure}
        \centerline{ \hbox{
               \epsfig{figure=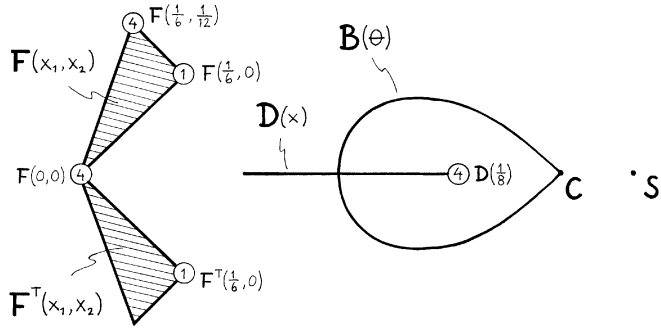,width=9cm}}}
        \caption{\small Here we show all MUB pairs that can be extended to 
triplets of MUBs, and how many triplets there are for each such pair, under 
the restriction that only 24th roots are used. The background for the picture 
is the set of all known equivalence classes of $N = 6$ Hadamard matrices.}
        \label{fig:2}
\end{figure}

\noindent Again this is all; these results are illustrated in fig.~\ref{fig:2}. 
In the picture the asymmetry between ${\bf F}(x_1,x_2)$ and 
${\bf F}^{\rm T}(x_1,x_2)$ looks quite odd. From section 3, and 
from eq. (\ref{FTF}) in particular, we know that the unordered 
MUB pairs $\{ {\id}, {\bf F}(1/6, 1/12)\}$ and $\{ {\id}, {\bf F}^{\rm T}
(1/6, 1/12)\}$ are equivalent, so they should have the same number of 
$(\mbox{MUB})_2$ candidates. In section 7 it will be seen that, in fact, they 
do---although in one case none of the candidates is built from 24th roots only.

We redid the calculation using 
48th, 72th, and 60th roots; $48 = 2^4\cdot 3$ and $72 = 2^3\cdot 3^2$ seemed 
like reasonable things to try, while $60 = 2^2\cdot 3\cdot 5$ was a wild shot. 
Anyway the list did not increase. It is tempting to think of vectors composed 
of 24th roots as forming a grid in a relevant part of Hilbert space. 
Through a study of the distances between the candidate $(\mbox{MUB})_2$ 
we could have refined the grid in interesting places, but we did not pursue 
this---the grid is not very dense. 

\vspace{1cm}

{\bf 6. MUB triplets including the Fourier matrix}

\vspace{5mm}

\noindent Let us consider the case when $(\mbox{MUB})_1$ is the Fourier matrix. 
Here we do not have to rely on our own calculations, because complete results are 
available: using the symbolic manipulation program MAGMA, 
Grassl \cite{Grassl} has computed all vectors unbiased with respect to both 
the standard and the Fourier basis. The same calculation was in fact done 
earlier by Bj\"orck and coworkers \cite{Froberg, Bjorck}. There are 48 such 
vectors altogether, or 24 vectors and their complex conjugates. Each vector 
shows up in two different bases, so they form 16 bases altogether. Using the 
classification of Hadamard matrices, the 16 $(\mbox{MUB})_2$ turn out to form 
four groups:

\

\noindent (i) Two circulant Fourier matrices enphased with 12th roots of unity; 
``$\omega {\bf F}$". \\
(ii) Two matrices equivalent to ${\bf F}^{\rm T}(1/6,0)$, and enphased with 
12th roots of unity; ``$\omega {\bf F}^{\rm T}$". \\
(iii) Six circulant Bj\"orck matrices ${\bf C}$ (three with $d$ and three with $\bar{d}$), 
enphased with 12th roots of unity; ``$\omega {\bf C}$". \\
(iv) Six Fourier matrices enphased with products of Bj\"orck's magical number $d$ 
(and $\bar{d}$) and 12th roots of unity; ``$d\omega {\bf F}$".

\

\noindent See eq. (\ref{d}) for the definition of $d$. We will refer to the 
bases in groups (i) and (ii) as classical, and to the remaining bases as 
non-classical, for a reason that will transpire. The matrices in groups (i) 
and (iii) are circulant when their columns are multiplied by suitable 
phase factors. They already include all vectors unbiased with the Fourier 
matrix. 
Both the $\omega {\bf F}^{\rm T}$ matrices contain three columns each from 
the two $\omega {\bf F}$ matrices. 
Each matrix $d\omega {\bf F}$ contains 
one vector from each matrix $\omega {\bf C}$, and vice versa. The four 
groups of bases are invariant under complex conjugation. Two of the bases in 
$\omega {\bf C}$, namely those given by Bj\"orck's matrix with $d$ and with 
$\bar{d}$, are unchanged by complex conjugation.

The distances between the 16 bases are as follows: The classical bases 
form a perfect square, with edge lengths squared $D_c^2 = 0.4$ (and the enphased 
Fouriers in opposite corners). The distance between any classical and any 
non-classical basis is always given by $D^2_c = 0.92$. The distance between 
any member of group (iii) to any member of group (iv) is always $D^2_c = 0.74$. 
Finally the distance relations 
within the two groups of 6 are identical; each group contains two 
equilateral triangles with edge lengths $D^2_c = 0.93$. These triangles are 
exchanged if $d$ and $\bar{d}$ are exchanged. In addition, within each group 
each basis has two bases at $D^2_c = 0.86$ and one at $D^2_c = 0.88$. 

\begin{figure}
        \centerline{ \hbox{
                \epsfig{figure=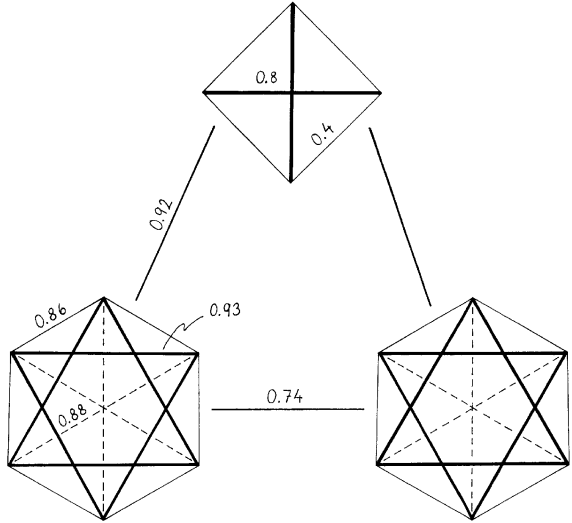,width=8cm}}}
        \caption{\small All Hadamard bases MUB with respect to ${\bf F}$ (a single 
point in fig.~\ref{fig:2}). The square consists of (enphased, affine) Fouriers, 
one of the two David's stars of Bj\"orck matrices 
enphased with 12th roots, and the other of Fouriers enphased with the number 
$d$. The numbers refer to the chordal distance squared, $D^2_c$.}
        \label{fig:3}
\end{figure}

The Bj\"orck/Grassl results are summarised in fig.~\ref{fig:3}. There does not 
exist any set of four MUBs including ${\id}$ and ${\bf F}$, since the distance 
squared between any possible pair of $(\mbox{MUB})_2$ never reaches 1. The best 
we can do with 7 bases is $D^2_c \geq 0.86$. This can be compared to random 
searches; in a sample of 20 million bases, chosen according to the measure 
described in section 4, the best results were $D^2_c \geq 0.91$ for 4 
bases, and $D^2_c \geq 0.86$ for 7 bases. 

It is striking that every vector 
vector unbiased with respect to ${\id}$ and ${\bf F}$ can be collected into some 
$(\mbox{MUB})_2$. The ``doubling'' of the non-classical bases is striking too.  
To explain these features we begin by 
proving that a vector is unbiased with respect to both ${\id}$ and ${\bf F}$ if 
and only if it is a member of a circulant Hadamard matrix. This requires us 
to recount some well known facts about the discrete Fourier transform. Given 
a sequence of complex numbers $z_a$, $0 \leq a \leq N-1$, we define its Fourier 
transform to be  

\begin{equation} \tilde{z}_a = \frac{1}{\sqrt{N}}\sum_{b=0}^{N-1}
q^{ab}z_b \ . \label{DFT} \end{equation}
  
\noindent In condensed notation 

\begin{equation} \tilde {z} = {\bf F}z \hspace{5mm} \Leftrightarrow \hspace{5mm} 
z = {\bf F}^{\dagger}\tilde{z} \ . \end{equation} 

\noindent If we are given a sequence of complex numbers $z_a$, the column vector 
whose components are $\tilde{z}_a/\sqrt{N}$ is unbiased with respect to the 
Fourier basis if and only if the sequence $z_a$ is unimodular, $|z_a|^2 = 1$, and 
it is unbiased with respect to the standard basis if and only if $\tilde{z}_a$ is 
unimodular. Hence vectors that are 
unbiased with respect to both the standard basis and the Fourier basis are 
in one-to-one correspondence to sequences obeying  

\begin{equation} |z_a|^2 = |\tilde{z}_a|^2 = 1 \end{equation}

\noindent for all values of $a$. Such sequences are called biunimodular. (Standard 
conventions for the Fourier transform sit a little uneasily with our conventions 
for MUBs. We decided to live with this, which is why our vectors are made from 
$\tilde{z}_a$ rather than $z_a$ itself.) 

Biunimodular sequences have an interesting property, that emerges when 
one studies the autocorrelation function 

\begin{equation} \gamma_b \equiv \frac{1}{N}\sum_{a=0}^{N-1}\bar{\tilde{z}}_a
\tilde{z}_{a+b} \ . \end{equation}

\noindent An easy calculation shows that 

\begin{equation} \gamma_b = \frac{1}{N}\sum_{a=0}^{N-1}
|z_a|^2q^{ab} \ . \end{equation}

\noindent Hence, if the sequence is biunimodular it obeys  

\begin{equation} \gamma_b = \delta_{b,0} \ . \end{equation} 

\noindent Therefore $\tilde{z}_a$ and $\tilde{z}_{a+b}$, with $b$ fixed and 
non-zero, are orthogonal vectors. 

A matrix is called circulant if it takes the form 

\begin{equation} C = 
\left[ \begin{array}{cccc} \tilde{z}_0 & \tilde{z}_{N-1} & \dots 
& \tilde{z}_1 \\ \tilde{z}_1 & \tilde{z}_0 & \dots & \tilde{z}_2 \\ \vdots & \vdots 
& \ & \vdots \\ \tilde{z}_{N-1} & \tilde{z}_{N-2} & \dots & \tilde{z}_0 
\end{array} \right] \ . 
\end{equation}

\noindent The matrix elements are 

\begin{equation} C_{ab} = \tilde{z}_{a-b} \ . \end{equation} 

\noindent With this definition a circulant matrix is Hadamard (up to 
normalisation) if and only if the sequence $z_a$ is 
biunimodular. It follows that all vectors unbiased with respect 
to both the standard and the Fourier bases can be collected into 
a set of circulant Hadamard matrices whose columns form bases that are 
MUB with respect to the standard and Fourier bases. 

This explains why all the 48 unbiased vectors are included in some basis 
that can be represented by an Hadamard matrix in circulant form. It 
remains to understand the ``doubling'' of bases found in Grassl's list. 
For any circulant matrix $C$, with first column $\tilde{z}_a$, it is easy 
to check that 

\begin{equation} {\bf F}^{\dagger}C = D{\bf F}^\dagger \ , \hspace{8mm} D = 
\mbox{diag}(z_0, z_1, \dots , z_{N-1}) \ . \label{CF} \end{equation}

\noindent Together with the earlier results, this entitles us to say that a 
vector is unbiased with respect to both the standard and the Fourier bases if 
and only if it is a column of a circulant Hadamard matrix, and every such matrix 
represents a basis that is MUB with respect to both ${\id}$ and ${\bf F}$. 
Incidentally it also implies that the Fourier matrix diagonalises every 
circulant matrix. Now consider an unordered MUB triplet $\{ {\id}, {\bf F}, 
H\}$. From section 3 we know that 

\begin{equation} \{ {\id}, {\bf F}, H\} \approx \{ {\bf F}^\dagger , {\id}, 
{\bf F}^\dagger H\} \approx \{ {\bf F}, {\id}, {\bf F}^\dagger H\} \ . 
\end{equation}

\noindent In the last step we multiplied ${\bf F}^\dagger$ from the right with 
$P \equiv {\bf F}^2$, which is a permutation matrix. 
Now let $H = C$, a circulant matrix. Using eq. (\ref{CF}) 

\begin{equation} \{ {\id}, {\bf F}, C\} \approx \{ {\bf F}, {\id}, 
{\bf F}^\dagger C\} \approx \{ {\bf F}, {\id}, D{\bf F}^{\dagger}\} 
\ . \end{equation}

\noindent This is how the second ``David's star'' of phased Fouriers, in 
fig.~\ref{fig:3}, arises. The transformation taking one David's star to 
another also exchanges the two enphased Fourier MUBs, while it leaves the 
pair in group (ii) invariant.  

For any dimension the conclusions are, first, that every vector unbiased 
with respect to both ${\id}$ and ${\bf F}$ is given by a biunimodular 
sequence, and conversely. Second, such a vector exists if and only if it 
is a member of a circulant basis that forms a MUB triplet with ${\id}$ 
and ${\bf F}$. Third, if this circulant basis 
is inequivalent (as an Hadamard matrix) to the Fourier matrix then there 
exists an equivalent MUB triplet whose third member is an enphased 
Fourier. We have no explanation for the two MUB triplets involving members 
of the ${\bf F}^{\rm T}$ family---from our point of view they arise by accident. 

Counting the number of unbiased vectors, or equivalently biunimodular sequences, 
remains a dimension dependent matter. By means of symbolic manipulation programs 
all biunimodular sequences with $N \leq 8$ entries have been listed by Bj\"orck 
and coworkers \cite{Froberg, Bjorck} (see also Haagerup \cite{Haagerup}). 
For $N = 6$ there are 48 biunimodular sequences (up to multiplication with a 
complex phase), or equivalently 48 vectors that are MUB with respect 
to both the standard and the Fourier bases. This includes 12 classical 
ones built from 12th roots---they were known to Gauss---and 36 non-classical 
ones built from Bj\"orck's magical number $d$.  

\vspace{1cm}

{\bf 7. MUB triplets with a block structure}

\vspace{5mm}

\noindent In section 5 we observed that there were several MUB triplets 
built from 24th roots. The ones that cannot be reduced to 12th roots are 
as follows:

\begin{itemize}
 
\item{Starting from the MUB pair $({\id}, {\bf D}(1/8))$ we have 4 triplets 
including enphased versions of ${\bf F}(1/6, 1/12)$. They form a 
semi-regular simplex where 4 edges have $D^2_c = 0.89$ and 2 edges have 
$D^2_c = 0.95$.} 

\item{Starting from the MUB pair $({\id}, {\bf F}(1/6, 1/12))$ we have 4 
triplets including enphased versions of ${\bf D}(1/8)$. They form a regular 
simplex where all the edges have $D^2_c = 0.93$.}  

\end{itemize}

\noindent The value $D^2_c = 0.95$ is the largest we have found in 
our searches for approximate MUB quartets.

These MUB triplets have an interesting structure in common. The affine 
family including the Fourier matrix, given in eq. (\ref{Fourier}), consists 
of Hadamard matrices that are equivalent under permutations to 

\begin{equation} {\bf F}(x_1,x_2) \approx {\bf F}_D \equiv 
\sqrt{\frac{3}{6}} \left[ 
\begin{array}{cc} {\bf F}_3 & {\bf F}_3 \\ {\bf F}_3D & - {\bf F}_3D 
\end{array} \right] \ , \hspace{12mm} D = \mbox{diag}(1, z_1, z_2) \ . 
\label{FD} \end{equation}

\noindent Here ${\bf F}_3$ is the 3 by 3 Fourier matrix. This is a kind of 
twisted version \cite{Haagerup,Dita} of the tensor product matrix 
${\bf F}_2\otimes {\bf F}_3$.
The Di\c{t}\u{a} family of matrices also has an interesting block structure. 
It was noted by Zauner that they can always be written on {\it block circulant 
form}, meaning that they can be brought to a form where the matrix consists 
of four blocks, each of which is a 3 by 3 circulant matrix \cite{Zauner}. 
This turns out to be relevant here. 

Consider the MUB pair $({\id}, {\bf F}(1/6, 1/12))$. Write it as 

\begin{equation} \left( {\id}, {\bf F}_D\right) \ , \hspace{10mm} 
D = \mbox{diag}(1, 1, \omega^6) \ , \hspace{8mm} \omega = 
e^{2\pi i/24} \ . \end{equation}
 
\noindent According to our list this pair can be extended to four triplets 
including an Hadamard matrix equivalent to ${\bf D}(1/8)$. With the 
members of the pair fixed according to the above, the four possible choices 
for $(\mbox{MUB})_2$ are 

\begin{eqnarray} \frac{1}{\sqrt{6}}\left[ \begin{array}{cc} C_1 & C_2 \\ 
C_2^{\dagger} & - C_1^\dagger \end{array} \right] \ , \hspace{8mm} 
\frac{1}{\sqrt{6}}\left[ \begin{array}{cc} C_2 & C_1 \\ 
C_1^{\dagger} & - C_2^\dagger \end{array} \right] \ , \nonumber \\ 
\ \\
\hspace{14mm}
\frac{1}{\sqrt{6}}\left[ \begin{array}{cc} C_1^\dagger & C_2^\dagger \\ 
C_2 & - C_1 \end{array} \right] \ , \hspace{8mm} 
\frac{1}{\sqrt{6}}\left[ \begin{array}{cc} C_2^\dagger & C_1^\dagger \\ 
C_1 & - C_2 \end{array} \right] \ , \nonumber \end{eqnarray}

\noindent where the $C_i$ are the 3 by 3 circulant matrices 

\begin{equation} C_1 = \left[ \begin{array}{ccc} 1 & \omega^{11} & \omega \\ 
\omega & 1 & \omega^{11} \\ \omega^{11} & \omega & 0 \end{array} \right] \ , 
\hspace{8mm} C_2 = \left[ \begin{array}{ccc} 1 & \omega^5 & \omega^7 \\ 
\omega^7 & 1 & \omega^5 \\ \omega^5 & \omega^7 & 1 \end{array} \right] 
\ . \end{equation}

\noindent In this sense the block structures of the Fourier and Di\c{t}\u{a} 
families are related. 

We now come to the following little puzzle. We know that 

\begin{eqnarray} \{ {\id}, \ {\bf F}(1/6, 1/12), \ {\bf D}(1/8) \} \approx 
\hspace{18mm} \nonumber \\ 
\approx \{ [{\bf F}(1/6, 1/12)]^\dagger, \ {\id}, \ [{\bf F}(1/6, 1/12)]^\dagger 
{\bf D}(1/8)\} \approx \\ 
\approx \{ {\id}, \ {\bf F}^{\rm T}(1/6, 1/12), \ DP[{\bf F}(1/6, 1/12)]^\dagger 
{\bf D}(1/8)\} \ , \hspace{2mm} \nonumber \end{eqnarray}

\noindent where $DP$ is the product of a diagonal unitary and a permutation matrix. 
Why then does ${\bf F}^{\rm T}(1/6, 1/12)$ not appear in fig.~\ref{fig:2}? 
The answer is that the matrix $[{\bf F}(1/6, 1/12)]^\dagger {\bf D}(1/8)$ is 
not built from 24th roots only.   

What is needed to resolve the puzzle is the number 

\begin{equation} b_1 = e^{2\pi i c_1} \ , \hspace{8mm} 
\cos{(2\pi c_1)} = \sqrt{\frac{2}{3}} \ . \end{equation}

\noindent We modified the ``24th roots program'' to allow also $b_1$ and 
$\bar{b}_1$ as entries in the vectors. We found the following new MUB triplets: 
 
\begin{itemize}

\item{Starting from the MUB pair $( {\id}, {\bf F}^{\rm T}(1/6, 1/12))$ we 
have 4 triplets including enphased versions of ${\bf F}^{\rm T}(c_1, 0)$.}

\item{Starting from the MUB pair $({\id}, {\bf F}^{\rm T}(c_1,0))$ we have 
2 triplets including enphased versions of ${\bf F}^{\rm T}(1/6, 1/12)$.}

\item{Starting from the MUB pair ${\bf F}(c_1,0)$ we have 2 triplets including 
enphased versions of ${\bf D}(-1/8)$.}

\end{itemize}

\noindent The first item in this list contains the ``missing'' triplets whose 
existence we already knew about---the asymmetries in fig.~\ref{fig:2} were due 
to the 24th roots restriction only. 

We have made computer searches to see if we can find other phase factors 
increasing the number of MUB triplets. One such number turned up, viz. 

\begin{equation} b_2 = e^{2\pi ic_2} \ , \hspace{8mm} 
\tan{(2\pi c_2)} = - 2 \ . \end{equation}

\noindent Again we modified the ``24th roots program'' to allow also $b_2$ and 
$\bar{b_2}$ as entries in the vectors. 
Then we found the following new MUB triplets:

\begin{itemize} 

\item{Starting from the MUB pair $({\id}, {\bf D}(0))$ we have 2 triplets 
including enphased versions of ${\bf F}(9/24 + c_2,0)$. The distance squared 
between the latter is $D^2_c = 0.77$.}  

\item{Starting from the MUB pair $({\id}, {\bf F}(9/24 + c_2,0))$ we have 120 vectors 
unbiased to both, of which 60 vectors form 10 bases 
that are enphased versions of ${\bf D}(0)$. They form a polytope where 9 edges end 
at each corner, 6 with $D^2_c = 0.93$ and 3 with $D^2_c = 0.78$.} 

\end{itemize}

\noindent So we have 6 bases with $D^2_c \geq 0.93$.

An explicit example of a MUB triplet including $b_2$ is 

\begin{equation} \left( {\id}, {\bf F}_D, {\bf D}_{\rm bc}\right) \ . \end{equation}

\noindent Here $D = \mbox{diag}(\omega^9b_2, 1, 1)$, ${\bf F}_D$ was defined in 
eq. (\ref{FD}), and ${\bf D}_{\rm bc}$ is a block circulant matrix equivalent to 
${\bf D}(0)$, namely 

\begin{equation} {\bf D}_{\rm bc} = \frac{1}{\sqrt{6}}
\left[ \begin{array}{cc} C_3 & C_4 \\ 
C_4 & -iC_3^\dagger \end{array} \right] \ , \end{equation}

\noindent where

\begin{equation} C_3 = \left[ \begin{array}{ccc} 1 & \omega^6 & \omega^6 \\ 
\omega^6 & 1 & \omega^6 \\ \omega^6 & \omega^6 & 1 \end{array} \right] \ , 
\hspace{8mm} C_4 = \left[ \begin{array}{ccc} \omega^{15} & \omega^3 & \omega^3 \\ 
\omega^3 & \omega^{15} & \omega^3 \\ \omega^3 & \omega^3 & \omega^{15} 
\end{array} \right] \ . \end{equation}

\noindent Thus it appears that many of the MUB triplets in this section can be 
brought to the (twisted product Fourier)---(block circulant Di\c{t}\u{a}) form. 
But we did not check them all, nor do we have any complete results 
analogous to those available for the $( {\id}, {\bf F})$ pair.

\vspace{1cm}

{\bf 8. What is the most general Hadamard matrix?}

\vspace{5mm}

\noindent We now take up the story of Hadamard matrices again. In section 
2 we gave a list of all known Hadamard matrices of order 6, or equivalently 
of all ordered MUB pairs up to equivalences. The question is whether this 
list is complete. To investigate this question one can try a 
perturbative approach to the unitarity equation: multiply the non-trivial 
matrix elements of a dephased Hadamard matrix with arbitrary phase factors, 
and solve the unitarity equation to first order in the phases. The number 
of free parameters that are left when this has been done is an integer 
known as the defect of the Hadamard matrix. It provides an upper bound on 
the dimensionality of any analytic set of Hadamard matrices \cite{Karol}. 

The defect for the known Hadamard matrices has been computed. For the Fourier, 
Bj\"orck, and Di\c{t}\u{a} matrices it equals 4, while it equals 0 for Tao's 
matrix. Closer inspection shows that the defect is likely 
to be constant within each affine family, with possible exceptions at isolated 
values of the affine parameters \cite{Karol}. We have also computed 
the defect for one thousand randomly chosen points along the non-affine 
family ${\bf B}(\theta)$; it was always equal to 4. These results show that 
Tao's matrix is an isolated point in the set of all Hadamard matrices. For 
the remaining matrices the answer is less clear. It appears that their defect 
is always equal to 4, but the defect provides us with 
an upper bound on the dimensionality only. It is not known whether the bound 
is attained, nor indeed whether additional Hadamard matrices, not connected 
to any of the above, exist. 

Let us be fully explicit about the calculation of the defect for 
Di\c{t}\u{a}'s matrix ${\bf D}$. We focus on 
this particular Hadamard matrix because its 
special form simplifies the resulting calculations. Starting from the known 
form of the matrix---see eq. (\ref{Dita2}), with $z = 1$---we multiply 
all non-trivial matrix elements with arbitrary phases according to  

\begin{equation} D_{ab} \rightarrow D_{ab}e^{ix_{ab}} \ , \hspace{5mm} 
1 \leq a, b \leq 5 \ . \label{defect} \end{equation}

\noindent The unitarity conditions give 15 complex equations, or 30 real 
equations; 
more than enough to determine the 25 parameters $x_{ab}$.  

We will try to solve the unitarity equations order by order in the phases. 
If we expand to first order in the phases we obtain the matrix 

\begin{equation} \frac{1}{\sqrt{6}}\left[ \begin{array}{cccccc} 
1 & 1 & 1 & 1 & 1 & 1 \\
1 & - 1 - ix_{11} & i - x_{12} & - i + x_{13} & - i + x_{14} & i - x_{15} \\
1 & i - x_{21} & - 1 - ix_{22} & i - x_{23} & - i + x_{24} & - i + x_{25} \\
1 & - i + x_{31} & i - x_{32} & - 1 - ix_{33} & i - x_{34} & - i + x_{35} \\ 
1 & - i + x_{41} & - i + x_{42} & i - x_{43} & - 1 - ix_{44} & i - x_{45} \\
1 & i - x_{51} & - i + x_{52} & - i + x_{53} & i - x_{54} & - 1 \end{array} 
\right] \ . \label{Dita} \end{equation}

\noindent Let us introduce the (unusual) notation $x_{aa}$, $x_{(ab)}$, 
$x_{[ab]}$ for the diagonal, symmetric off-diagonal, and anti-symmetric 
parts of $x_{ab}$. One finds that the 30 equations split naturally into

\

5 equations for $x_{aa}$

10 equations for $x_{(ab)}$ 

15 equations that split into 6 for $x_{[ab]}$ and 9 for $x_{(ab)}$.
 
\

\noindent To first order all the equations are linear. There will be 4 
undetermined parameters in $x_{[ab]}$. One also finds  

\begin{equation} x_{aa} = 0 \hspace{12mm} x_{(ab)} = 0 \ . \end{equation}

\noindent Since the 19 equations for $x_{(ab)}$ are linear there are no consistency 
problems. It is however worth observing that the general solution of the final 
9 equations for $x_{(ab)}$ is 

\begin{equation} x_{(ab)} = x \ , \end{equation}

\noindent with $x$ remaining as a free parameter, to be set to zero 
by the other 10 equations for $x_{(ab)}$. 

A suitable choice for the free parameters in $x_{[ab]}$ is  
$x_{[12]}$, $x_{[13]}$, $x_{[24]}$, $x_{[34]}$.
Then one finds

\begin{equation} \begin{array}{l} 
x_{[14]} = x_{[12]} + x_{[34]} \\
x_{[15]} = x_{[13]} + x_{[34]} \\
x_{[23]} = x_{[13]} + x_{[24]} \\
x_{[25]} = - x_{[12]} + x_{[13]} \\
x_{[35]} = - x_{[24]} + x_{[34]} \\
x_{[45]} = - x_{[12]} - x_{[24]} 
\ . \end{array} \end{equation}

\noindent In conclusion, to first order there are 4 free phases, or equivalently 
the defect equals 4. 

We have carried the calculation to third order. However, the Ansatz in eq. 
(\ref{defect}) must then be modified. Set 
 
\begin{eqnarray} D_{ab} \rightarrow D_{ab}e^{i(x_{ab} + y_{ab} + z_{ab} + 
\dots )} = \hspace{58mm} \nonumber \\
\ \label{expansion} \\
= D_{ab}\left( 1 - \frac{1}{2}x_{ab}^2 - x_{ab}y_{ab} + 
i(x_{ab} + y_{ab} + z_{ab} - \frac{1}{6}x_{ab}^3) + \dots \right) \ . 
\nonumber \end{eqnarray}

\noindent Here it is understood that $y_{ab}$ and $z_{ab}$ are quadratic and 
cubic functions of $x_{ab}$, respectively. For obvious reasons we do not write 
out the resulting matrix here. There is an amount of ambiguity in the 
definition of the higher order terms. This we resolve by imposing  

\begin{equation} y_{[12]} = y_{[13]} = y_{[24]} = y_{[34]} = 0 \ , 
\label{condition} \end{equation}

\noindent and analogously at all higher orders. This can always be achieved 
through a redefinition of the free parameters in $x_{ab}$. Once this is done 
the expansion in the exponent of eq. (\ref{expansion}) is unambiguous. 

At each order higher than the first, there will be 30 linear inhomogeneous 
equations for the $25 - 4$ new parameters that we introduce, with the inhomogeneous 
terms made up from products of known lower order contributions. Given eq. 
(\ref{condition}), the anti-symmetric and diagonal parts will be completely 
determined. There remain 10 + 9 equations for the symmetric off-diagonal 
part, and it is not {\it a priori} clear that a solution exists. 

At second order one finds 

\begin{equation} y_{[ab]} = 0 \end{equation}

\begin{equation} \begin{array}{l} 
2y_{11} = - x_{21}^2 + x_{31}^2 + x_{41}^2 - x_{51}^2 \\
2y_{22} = - x_{12}^2 - x_{32}^2 + x_{42}^2 + x_{52}^2 \\
2y_{33} = x_{13}^2 - x_{23}^2 - x_{43}^2 + x_{53}^2 \\
2y_{44} = x_{14}^2 + x_{24}^2 - x_{34}^2 - x_{54}^2 \\
2y_{55} = - x_{15}^2 + x_{25}^2 + x_{35}^2 - x_{45}^2 \ . 
\end{array}  \end{equation}

\noindent The last 9 equations imply 

\begin{equation} y_{(ab)} = y \ , \label{y} \end{equation}

\noindent for some $y$. Detailed examination shows that the remaining 
10 equations are consistent with this and with each other. The 
solution can be written as  

\begin{equation} 4y = 2y_{11} + 2y_{22} - (x_{13} - x_{23})^2 
+ (x_{14} - x_{24})^2 - (x_{15} - x_{25})^2 \ . \end{equation} 

\noindent The expansion to second order is therefore consistent. 

At third order one finds 

\begin{equation} z_{aa} = 0 \ , \end{equation}

\noindent together with rather complicated expressions for the part of 
$z_{[ab]}$ left undetermined by the condition analogous to 
(\ref{condition}). We do not give them here. The first 10 equations for 
the symmetric off-diagonal part is solved by 

\begin{equation} z_{(ab)} = 0 \ . \end{equation}

\noindent Thus $z_{ab}$ is, like $x_{ab}$, anti-symmetric in its indices. 
We are then left with 9 consistency conditions on this solution. Again 
detailed examination shows that they are obeyed, hence the expansion is 
consistent to third order. 

In conclusion we have proved that Di\c{t}\u{a}'s matrix admits an expansion 
to third order in four arbitrary phases. Unfortunately, to second and third 
order this result depends on cancellations that we do not understand---we 
have simply checked that they do occur. Nevertheless it seems to us likely that 
the expansion exists to all orders. This suggests that the parameter space of 
allowed Hadamard matrices has a four dimensional component. 

Some solutions are known to all orders. First of all, if the expression 
$D_{ab} \rightarrow D_{ab}e^{ix_{ab}}$ is exact, that is if all higher order 
terms in the exponent vanish, we have an affine family. It is known that 
Di\c{t}\u{a}'s matrix admits 5 permutation equivalent affine families 
\cite{Karol}. They result from the five choices 

\begin{equation} \begin{array}{llll} 
\mbox{i:} & \ & (x_{[12]}, x_{[13]}, x_{[24]}, x_{[34]}) = (x,0,0,0) &  \\
\mbox{ii:} & \ & (x_{[12]}, x_{[13]}, x_{[24]}, x_{[34]}) = (0,x,0,0) &   \\
\mbox{iii:} & \ & (x_{[12]}, x_{[13]}, x_{[24]}, x_{[34]}) = (0,0,x,0) &  \\
\mbox{iv:} & \ & (x_{[12]}, x_{[13]}, x_{[24]}, x_{[34]}) = (0,0,0,x) &  \\
\mbox{v:} & \ & (x_{[12]}, x_{[13]}, x_{[24]}, x_{[34]}) = (x,x,-x,-x) & . 
\end{array}  \end{equation}
 
\noindent In fact our choice of independent variables was made with this 
result in view. In these cases we have $y_{ab} = z_{ab} = 0$, and similarly 
for all higher order terms, so these solutions are somewhat trivial. 
Fortunately at least one non-affine family, with non-vanishing higher order 
terms, is also known to all orders. This is the one-parameter family 
${\bf B}(\theta)$, which passes through representatives of the Di\c{t}\u{a} 
equivalence class no less than three times. A solution to our equations, 
equivalent to ${\bf B}(\theta)$ to third order, is 

\begin{equation} (x_{[12]}, x_{[13]}, x_{[24]}, x_{[34]}) = 
(-x, -x, 2x, x) \ . \end{equation}

\noindent The higher order terms in the exponent of eq. (\ref{expansion}) are 
now non-vanishing. At the same time we observe that, even if it does exist to 
all orders, this solution must encounter convergence problems in order to 
reproduce the limits on the phase $\theta$ in the Beauchamp-Nicoara family. 
Clearly such a problem can occur in the exponent of eq. (\ref{expansion}). 

Our calculation, together with several different one-parameter solutions known 
to all orders (one of which is a non-affine solution), does support the 
conjecture that the set of Hadamard matrices is four dimensional in a 
neighbourhood of the Di\c{t}\u{a} matrix. We understand that Beauchamp and 
Nicoara have numerical evidence for the same state of affairs in a neighbourhood 
of the Bj\"orck matrix \cite{Remus}. At the same time the existence of Tao's 
matrix shows that the set of Hadamard matrices must be disconnected. The Bj\"orck 
and Di\c{t}\u{a} matrices are connected to each other, but we have 
no evidence to show how the Fourier matrix fits into the picture.  

\vspace{1cm}

{\bf 9. Summary}

\vspace{5mm}

\noindent Mutually unbiased bases live in the landscape of complex Hadamard 
matrices. A complete classification of the equivalence classes of $N = 6$ 
complex Hadamard matrices, or equivalently of the set of all MUB pairs, 
is however not available. If we insert arbitrary 
phases in the known examples, we find that four parameters are left 
undetermined to first order. The one exception is Tao's matrix. For 
Di\c{t}\u{a}'s matrix we solved the unitarity equations to third order in 
four arbitrary phases. This does not constitute a proof that the parameter 
space has a four dimensional component, but we may perhaps quote 
Lewis Carroll at this point: ``I have told you thrice. What I tell you 
three times is true.'' 

A set of MUBs can always be described as the standard basis together with 
a set of Hadamard matrices. We searched for such sets, mostly under the 
restriction that all vectors be composed from rational roots of unity, 
but we also gave three examples of non-rational and ``MUB-friendly'' phase 
factors. The restriction to rational roots is severe, but the advantage is 
that systematic searches can be made.  

We found only triplets of MUBs. A 
rich source of triplets is obtained by choosing the Fourier basis as one of 
the members; the resulting structure can be largely understood through the 
discrete Fourier transform. Other triplets, with a special block structure, 
were also found. 
To add structure to the problem we introduced a natural distance between 
bases in Hilbert space. It attains its maximum value when the bases are 
MUB. Whenever a pair of MUBs (including 
the standard basis) could be extended to a triplet in more than one way, we 
computed the distance between the possible third members. In this way we 
found sets of 4 bases where all distances 
but one are maximal. The highest figure we obtained for the last distance was 
$D^2_c = 0.95$, significantly closer to the maximum value 1 than is the 
average distance squared between bases, $\langle D^2_c\rangle = 6/7$. 

No-go statements for complete sets of 7 MUBs in $N = 6$ have been made. They 
tend to exclude sets with a high degree of symmetry \cite{KP} or with otherwise 
special group theoretical properties \cite{Aschbacher}, as well as sets constructed 
using generalisations of methods that work in prime power dimensions \cite{Archer}. 
We are unsurprised by our failure to find any 
complete set---but if our conjecture concerning the set of all Hadamard matrices 
is true, the no-go theorems are unlikely to apply in general. 
  
So what is the best approximation to a complete set of MUBs that one can 
find in 6 dimensions? Indeed, what is the maximal number of MUBs one 
can have in 6 dimensions? The jury is still out.  

Note added in proof: After submission of this paper Matolcsi and Sz\"ollosi \cite{Matolcsi} found a new 
family of 6   6 Hadamard matrices. And in Ref. \cite{Tadej}, Tadej, \.Zyczkowski, and Slomczynski study 
the defect of unitaries, not necessarily Hadamard.

\vspace{8mm}

{\bf Acknowledgements:}

\vspace{4mm}

\noindent This work grew out of the Master's Thesis of one of us (WB). 
We thank Kyle Beauchamp and Remus Nicoara for sharing their results 
with us prior to publication, Markus Grassl for sending his vectors for 
inspection, Bengt Nagel for drawing our attention to the book by Kostrikin 
and Tiep, and S\"oren Holst for helping us to the result described in 
fig.~\ref{fig:1}. 

We acknowledge partial financial support from the Swedish Research Council, 
from grant 
PBZ-MIN-008/P03/2003 of the Polish Ministry of Science and Information 
Technology, and from the European Union research project SCALA.

\end{document}